\documentclass{emulateapj}
\usepackage{amssymb}
\usepackage{lscape}
\usepackage{longtable}
\newcommand{\minpoint}{\mbox{$'\mskip-4.7mu.\mskip0.8mu$}}

\newcommand{\subsun}{\mbox{$_{\odot}$}}
\newcommand{\et}{{\rm et al.}~}

\def\ltsima{$\; \buildrel < \over \sim \;$}
\def\simlt{\lower.5ex\hbox{\ltsima}}
\def\gtsima{$\; \buildrel > \over \sim \;$}
\def\simgt{\lower.5ex\hbox{\gtsima}}

\def\arcm{$'~$}

\begin{document}
\title{SPECTROSCOPIC IDENTIFICATION OF A PROTO-CLUSTER AT ${\rm z=2.300}$: 
ENVIRONMENTAL DEPENDENCE OF GALAXY PROPERTIES AT HIGH REDSHIFT}  

\slugcomment{DRAFT: \today}
\author{\sc Charles C. Steidel}
\affil{California Institute of Technology, MS 105-24, Pasadena, CA 91125}
\author{\sc Kurt L. Adelberger\altaffilmark{2}}
\affil{Observatories of the Carnegie Institution of Washington, 813 Santa Barbara Street,
Pasadena, CA 91101} 
\author{\sc Alice E. Shapley\altaffilmark{3}}
\affil{University of California, Berkeley, Department of Astronomy, 601 Campbell Hall,
Berkeley, CA 94720}
\author{\sc Dawn K. Erb and Naveen A. Reddy}
\affil{California Institute of Technology, MS 105-24, Pasadena, CA 91125}
\author{\sc Max Pettini}
\affil{Institute of Astronomy, Madingley Road, Cambridge CB3 OHA UK}

\altaffiltext{1}{Based, in part, on data obtained at the 
W.M. Keck Observatory, which 
is operated as a scientific partnership among the California Institute of 
Technology, the
University of California, and NASA, and was made possible by the generous 
financial
support of the W.M. Keck Foundation. Also based on data obtained during the in-orbit
check-out of the {\it Spitzer Space Telescope}.}
\altaffiltext{2}{Carnegie Fellow}
\altaffiltext{3}{Miller Fellow}
\begin{abstract}
We have discovered a highly significant over-density of galaxies at
$z=2.300\pm0.015$ in the course of a redshift survey designed to
select star-forming galaxies in the redshift range $z=2.3\pm0.4$ 
in the field of the bright $z=2.72$ 
QSO HS1700+643. The structure has a redshift-space
galaxy over-density of $\delta_g^z\simeq7$ and an estimated matter over-density
in real space of $\delta_m \simeq 1.8$, indicating that it will virialize
by $z\sim 0$ with a mass scale of $\simeq 1.4\times 10^{15}$ M$_{\sun}$, that of 
a rich galaxy cluster. Detailed modeling of the
spectral energy distribution-- from the rest-far-UV to the rest-near-IR -- of 
the 72 spectroscopically confirmed galaxies in this field for which
we have obtained $K_s$ 
and {\it Spitzer}/IRAC photometry, 
allows for a first direct comparison of galaxy
properties as a function of large-scale environment at high redshift. We find
that galaxies in the proto-cluster environment have mean stellar masses and inferred
ages that are $\sim 2$ times larger (at $z=2.30$) than identically UV-selected
galaxies outside of the structure, and show that this is consistent with
simple theoretical expectations for the acceleration of structure formation
in a region that is over-dense on large scales by the observed amount. The proto-cluster
environment contains a significant number of galaxies that already appear
old, with large stellar masses ($>10^{11}$ M$_{\sun}$), by $z=2.3$.   
\end{abstract}
\keywords{cosmology: observations --- galaxies: evolution --- galaxies: high-redshift --- 
galaxies: clusters}

\section{Introduction}
\label{sec:intro}

The process of galaxy formation is almost universally acknowledged to be
dependent on large-scale environment, with the regions of the universe
that are most dense on large scales expected to be the sites of the earliest
galaxy formation.  In the present-day universe, regions that were over-dense
on $\sim 10$ Mpc scales at high redshift have virialized to become rich galaxy clusters, harboring
much more than their fair share of galaxies with the largest stellar 
masses and (apparently) the oldest luminosity-weighted stellar ages (e.g., \citealt{hogg2003}).
Measurements of galaxy clustering statistics at low (e.g., \citealt{budavari2003}),
intermediate (e.g., \citealt{coil2004}) and high (e.g.,
\citealt{daddi2003,adelberger2004b,adelberger2005})
redshift all indicate highly significant differences in clustering amplitude as a function
of galaxy color and/or rest-frame optical luminosity, in the sense that redder (older)
and  more optically luminous (larger stellar mass) galaxies are more strongly clustered.
There has been a vast literature over the past $\sim 30$ years 
focused on the properties of early type galaxies, both in clusters (e.g., \citealt{faber1973,bower1992,
ellis1997,stanford1998,vandokkum2001,holden2004}) at low and intermediate redshifts and in the
``field'' (e.g., \citealt{treu1999,kochanek2000,vandokkum2003}), attempting to establish
the formation epochs of these galaxies and the differences, if any, as a function of
environment (e.g., \citealt{thomas2004}). Virtually all of these studies find that 
early type galaxies must have formed most of their stars early ($z > 2$) and that they
have evolved largely passively since that time. There is less agreement about the extent
to which the galaxies' history depends on environment, and what low or intermediate
redshift results imply about the scatter in ages within clusters or the difference
between clusters and lower density environments.  One of the issues that undoubtedly
clouds the debate is what constitutes an ``early-type'' galaxy or even a ``cluster''. 
Most of the galaxies studied have been selected to have the morphological
appearance of early type galaxies, possibly leading to different subsets of galaxies
being selected at different epochs (called ``progenitor bias'' by \citealt{vandokkum2000}).

Very recently, considerable attention has been directed to photometric and
spectroscopic surveys of galaxies at high redshifts ($z\simlt 2$) that are bright
and/or red at rest-frame optical wavelengths (e.g., \citealt{franx2003,abraham2004,
cimatti2004,mccarthy2004}), as might be expected for passively evolving galaxies
at high redshifts. A general conclusion seems to be that passively evolving galaxies
with large stellar masses ($M^{\ast} > 10^{11}$ M$_{\sun}$) are present out to
$z \sim 2$ with a number density roughly similar to that observed locally. The inferred
star formation histories for these galaxies imply that most of the stars 
formed prior to $z \sim 3$, in good agreement with inferences based
on the evolution in the fundamental plane of cluster ellipticals to $z \simgt 1$ (e.g.,
\citealt{holden2004}). However, the number
of spectroscopically identified massive galaxies in these new spectroscopic surveys
is far too small to establish their large-scale context.   

At present, the only spectroscopic samples of galaxies at $z \simeq 2$ that are
large enough for a direct comparison of galaxy properties as a function of
environment are those that are selected by color in the rest-UV (e.g., \citealt{steidel2003,
steidel2004}).  Obviously, such samples depend on galaxies 
having enough current (relatively unobscured)
star formation to be identified as a candidate for spectroscopy, which will produce a
bias against both passively evolving $z>2$ galaxies and galaxies that are heavily
reddened by dust (cf. \citealt{daddi2004}). However, the rest-UV-selected surveys do 
turn up substantial numbers of galaxies with $M_{\ast} > 10^{11}$ M$_{\sun}$ whose physical
properties are plausibly consistent with those 
observed in the rest-optically selected surveys \citep{shapley2004,shapley2005,adelberger2005}.  
\citet{adelberger2004b} have argued, based on their number density and clustering
properties, that the galaxies present in the spectroscopic
UV-selected samples at $z \sim 2-3$ are likely to be the general progenitors of 
what would be called ``red'' galaxies at $z \sim 1$ and ``early-type'' galaxies by
$z \sim 0.2$, independent of environment.  
Moreover, the surveys are tuned to be sensitive to star-forming galaxies during 
the epoch when the bulk of star formation in massive early-type cluster galaxies is inferred  
to have occurred. With the advent of the {\it Spitzer Space Telescope} and its capability
of obtaining high-quality photometry in the rest-frame near-IR at $z \sim 2-3$ using
the IRAC instrument \citep{fazio2004}, it is now possible to obtain relatively reliable
measures of stellar population parameters, particularly stellar masses, when combined
with information available from large ground-based telescopes. 
 
In this paper, we analyze results from a spectroscopic survey
in which we have discovered a significant redshift-space over-density of galaxies 
at $z=2.30$. While the characterization of galaxy over-densities at high redshift is 
being pursued using similar spectroscopic samples (e.g., \citealt{steidel1998})
and through targeted observations of likely proto-cluster regions marked by high
redshift radio galaxies (e.g., \citealt{kurk2000,venemans2002}) or multiple sub-mm-selected
galaxies (\citealt{blain2004}), the combination of spectroscopic and photometric data available in
the field of QSO HS1700+643 (notably, the availability of {\it Spitzer} data) 
provides an unprecedented view of the properties of both the
proto-cluster galaxies and their larger-scale context at high redshift: 
because the selection function of our spectroscopic sample is broad and is roughly
centered on the redshift of the over-density, we are able to directly compare the overall
properties of similarly-selected galaxies within and outside of a structure that (we
argue) will become a rich galaxy cluster by the present epoch.  The relative insensitivity
of our rest-UV selection criteria to stellar mass or star formation age
offers a distinct advantage in this case: the spectroscopic sample, covering the redshift range $2.29\pm0.32$
(mean and standard deviation) in the HS1700+643 field, offers a relatively unbiased view
of the progress of galaxy evolution within and outside of the large structure. 
We show, using the results of detailed modeling presented in \citet{shapley2005}, that
there are already clear differences in the progress of galaxy evolution between the
proto-cluster and ``field'' environment at $z \sim 2.3$, when the Universe was 2.8 Gyr
old. Large scale environment, and not redshift, is likely to be the primary factor
determining the maturity of massive galaxies in the distant universe.  

We assume a cosmology with $\Omega_m=0.3$, $\Omega_{\Lambda}=0.7$, and $h=0.7$ throughout.

\section {Observations and the Galaxy Sample}
\label{sec:obs}

\begin{figure}
\plotone{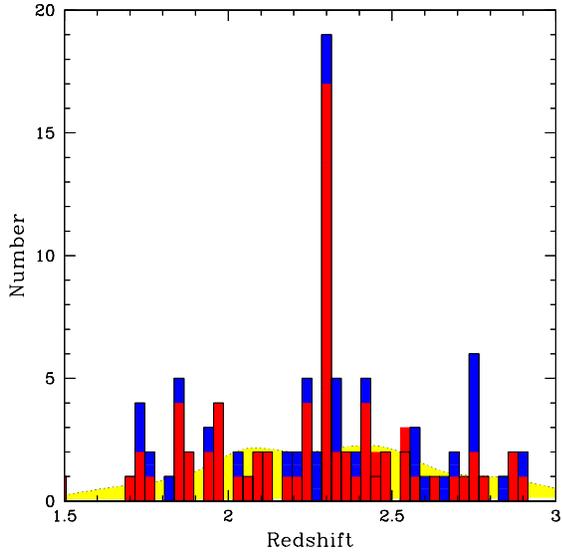}
\caption{
Redshift histogram for spectroscopically identified galaxies in the HS1700$+$643 field. The blue
(background) histogram shows the redshift distribution for the 100 spectroscopic redshifts in
the field; the red (foreground) histogram shows the redshift distribution of the 72 objects
with IRAC measurements. The light shaded curve in the background shows the expected redshift
distribution of a sample of 72 objects selected
from a combination of ``BX'' and ``MD'' candidates, in the observed
ratio of $\sim 4.5:1$.
}
\label{fig:zhist}
\end{figure}

\begin{figure}
\plotone{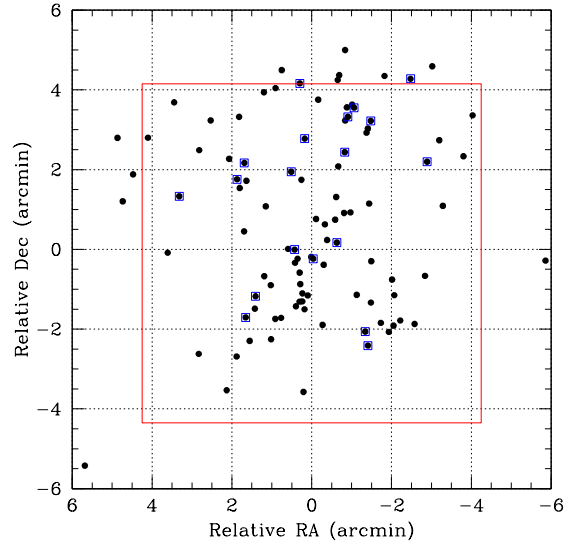}
\caption{
A map on the plane of the sky of galaxies with redshifts shown in figure \ref{fig:zhist}.
The coordinates are arcmin relative to the position of the background QSO HS1700+643, and
the (red) square represents the approximate region covered by the ground-based $K_s$
band images, most of which is also covered by the IRAC images (see figure 1 of \citet{shapley2005}).
Galaxies with $z=2.300\pm0.015$ are indicated with open (blue) boxes. The distribution of
spectroscopically observed galaxies is highly non-uniform due to the slitmask geometry of
LRIS; within the region bounded by the 8.5 arcmin square, 24\% of the galaxies that satisfy
the BX/MD color criteria and are detected in the $K_s$ band to $K_s\simeq 22.0$ (Vega)
have spectroscopic IDs.}
\label{fig:field_map}
\end{figure}

The spectroscopic observations in the HS1700+643 field were obtained as part of a
survey designed to simultaneously explore both star-forming galaxies at $z \sim 2-2.5$
and the interaction between the galaxies and the intergalactic medium (IGM) in the
same volumes of space.  The initial results of the general $z \sim 2$ spectroscopic survey are
described by \citet{steidel2004}, and the full results of the sub-sample in the HS1700
field, including a description of all of the photometric observations, 
are presented by \cite{shapley2005}. Because the redshift of HS1700+643 is
$z_Q=2.72$, we used optical photometric criteria that would sample galaxies with
$z_{\rm gal} \simeq 2-2.75$, i.e., foreground to the QSO; thus we used a combination
of the ``BX'' criteria of \citet{adelberger2004a} ($z=2.20\pm0.32$) and the ``MD'' criteria
of \citet{steidel2003} ($z=2.73\pm 0.27$). Accounting for the 4.5:1 ratio in the number 
of spectroscopically identified BX and MD candidates to ${\cal R} = 25.5$ (the actual
ratio of photometric candidates is 6:1 to the same apparent magnitude limit), 
the expected redshift distribution for a random spectroscopic sample would be $\langle z
\rangle= 2.26\pm0.35$, 
based on the overall observed redshift selection functions of BX and MD 
candidates in all survey fields.  A total of 100 spectroscopic redshifts in the
range $1.48 \le z \le 2.89$ has been obtained in this field to date, with the redshift
distribution (shown in figure \ref{fig:zhist})
of $\langle z \rangle =2.28\pm0.32$ (mean and standard deviation), 
entirely consistent with the expectations 
based on the overall redshift selection functions of the BX/MD candidates. The distribution
of observed galaxies on the plane of the sky is strongly concentrated within $\sim 4-5$ arcmin of
the line of sight to HS1700+64, as shown in figure \ref{fig:field_map}, largely dictated
by geometric constraints of the LRIS slitmasks used for the spectroscopic observations
given our emphasis on galaxies near the QSO sightline. 

The galaxies with spectroscopic redshifts represent a very small fraction of the photometric
candidates in the field, which total 1711 BX+MD objects to ${\cal R}=25.5$ over the full
15\arcm\ by 15\arcm\ field. Within the region of the field that has also been observed in
the $K_s$ band (figure \ref{fig:field_map}), the fraction of candidates with spectroscopic
identifications is
$\sim 19$\% (98/508, of which 84 have $z>1.4$\footnote{Within the same region, a total of 
162 objects was observed spectroscopically,
so that 60.5\% of objects attempted yielded redshifts. This fraction is
very close to the overall fraction of ``BX'' candidates that yield redshifts in the full spectroscopic sample. 
As discussed by \citet{steidel2004}, failures to measure redshifts result from inadequate S/N after
90 minutes of integration (often dominated by masks observed under marginal conditions).
Deeper spectroscopic observations show that the initial failures have the same redshift distribution as those that 
are successful on a first pass.}   
and the fraction of $K_s$-detected candidates with redshifts is
$\sim 24$\% (92/389, of which 79 have $z > 1.4$).  
Because the remainder of the discussion of the galaxies in the field
depends on the longer wavelength $K_s$ and IRAC data, hereinafter we focus only on
the sample of 72 $z > 1.4$ galaxies with spectroscopic redshifts that are also detected
in both $K_s$ and at least one IRAC band (see \citet{shapley2005}). The redshift
distribution for this sub-sample is shown in figure \ref{fig:zhist} as the red histogram
in the foreground. It is important to note that the galaxies were chosen for
spectroscopic observation without regard to anything 
but position relative to the QSO sightline and
optical magnitudes and colors. Also shown in figure \ref{fig:zhist} is the redshift
distribution expected for a sample of 72 galaxies drawn from a BX+MD-selected sample
with the observed BX/MD ratio. This distribution, which is that expected in the absence
of clustering, will be used to evaluate the significance of the over-density in \S\ref{sec:results}.

All of the photometric measurements and inferred stellar population parameters 
for the 72 galaxies in the sample are taken directly
from \citet{shapley2005}.

\section{Results}
\label{sec:results}

\subsection{Quantifying the Over-density}
\label{sec:overdense}

As shown in figure \ref{fig:zhist}, there is an obvious galaxy concentration
at $z\simeq 2.30$ in the HS1700$+$64 field, evident both in the subset
of galaxies in the $K_s$/IRAC observed region of the field and in the larger 
optical field.  
We estimate the significance of the spike using
the method described in Steidel et al. (1998).
Briefly, we consider every pair of galaxies in turn,
calculate the redshift difference between them and
count the number of additional galaxies whose redshifts
lie between the redshifts of the pair.  Regions where a large
number of galaxies appear in a small redshift interval are identified
as possible clusters.  The strength of each cluster is quantified
with a statistic, $\zeta$ (see \citealt{steidel1998}).
We then focus our attention on the cluster candidate with the 
smallest value of $\zeta$ (i.e., with redshifts that are most inconsistent
with a random distribution), and calculate the significance of the
cluster by analyzing numerous randomized galaxy catalogs in the
same way and counting how frequently a value of $\zeta$ so small (or smaller)
appears.

\begin{figure*}
\plottwo{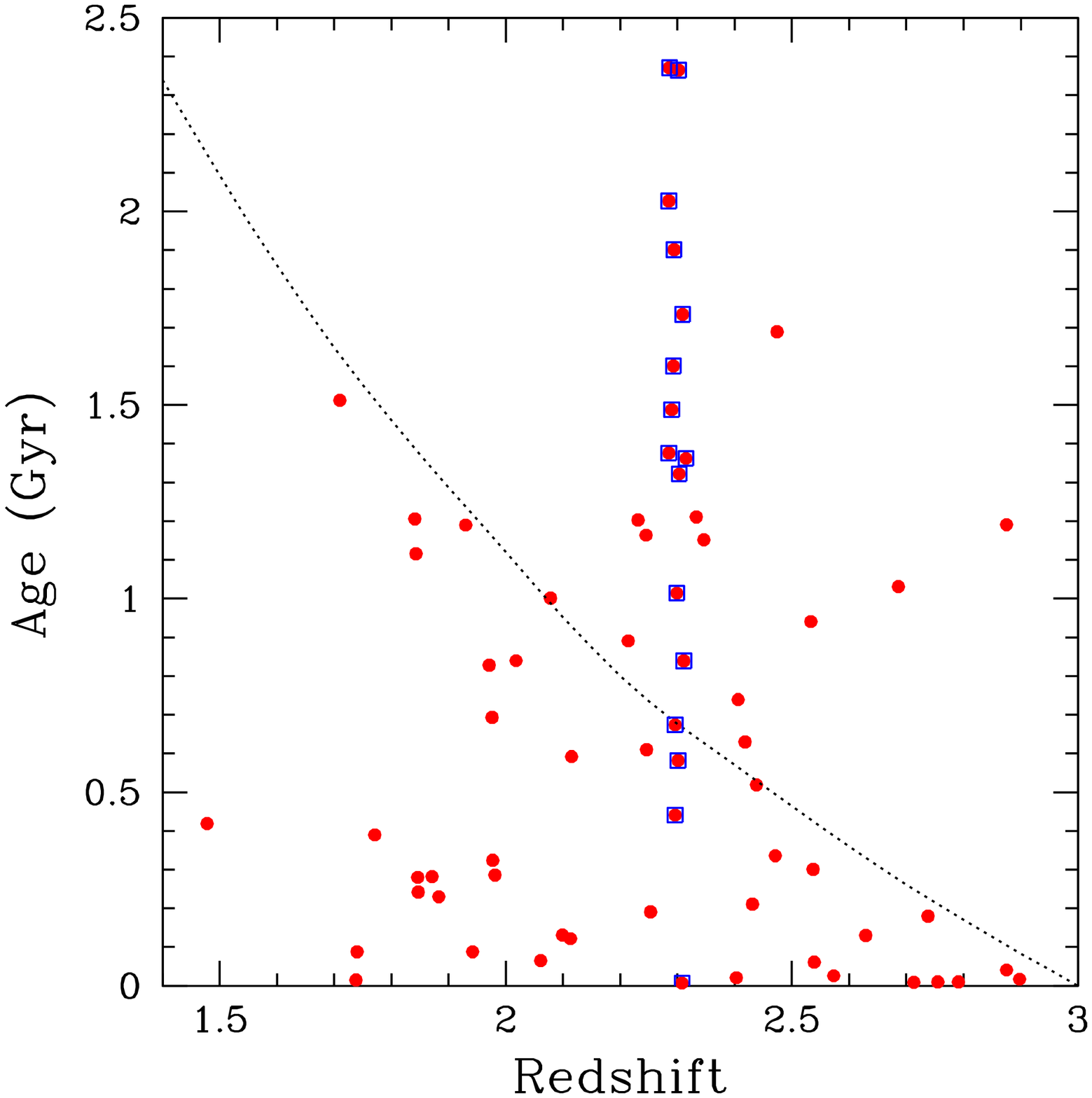}{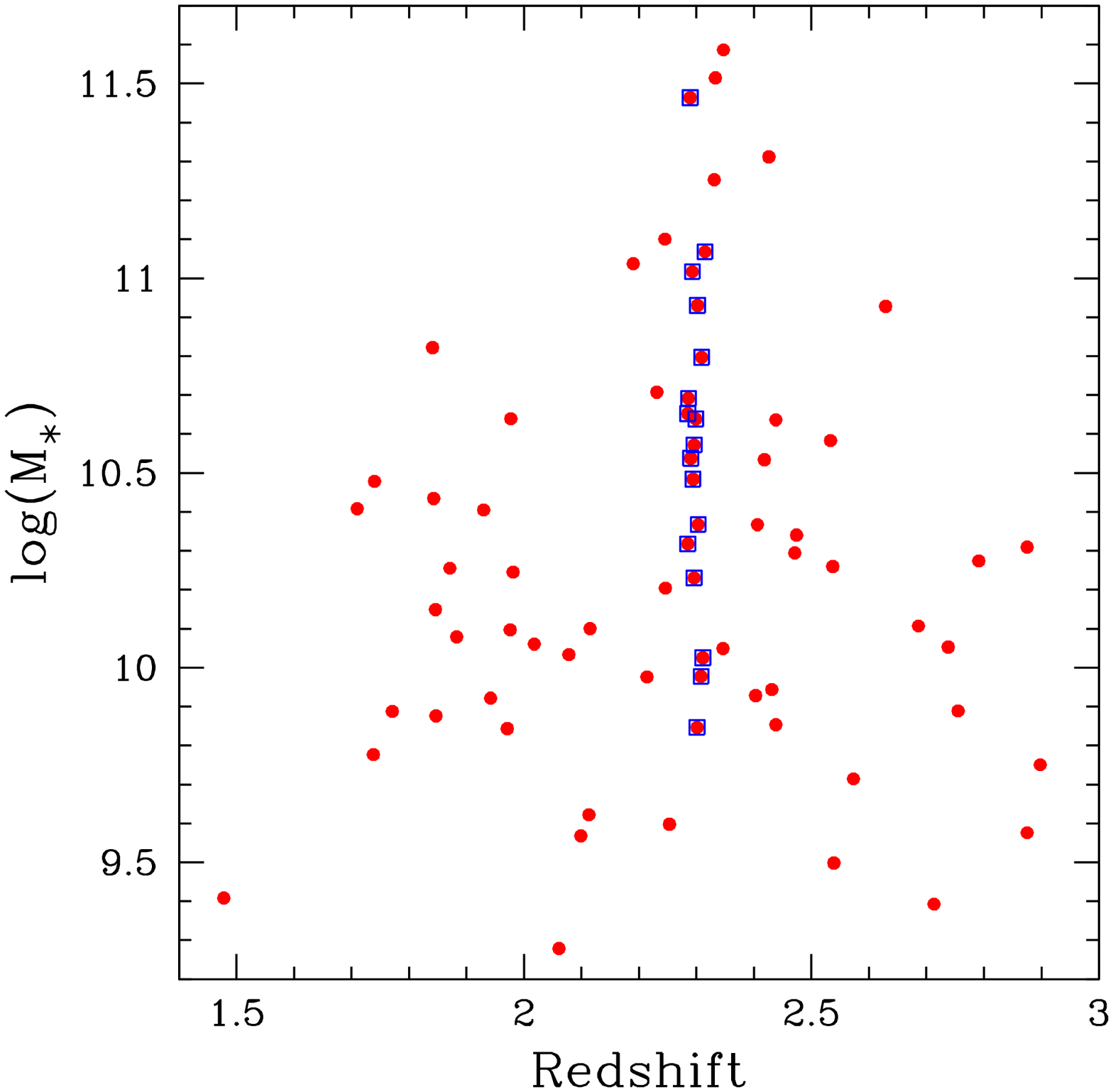}
\caption{The inferred age for constant star formation (CSF) models (left) and stellar mass (right) versus redshift for the 72 galaxies in the
sample. Note that, on average, objects within the structure at $z=2.300\pm0.015$ are evidently both older
and more massive than galaxies outside of it. The dotted line in the left-hand panel indicates
age as a function of redshift for a galaxy which began forming stars at $z=3.0$. 
}
\label{fig:mass_age_z}
\end{figure*} 

This approach isolates the region between
$z=2.285$ and $z=2.315$ (i.e., $z=2.30\pm0.015$) as the most significant over-density
in our data, containing 17 galaxies within the region observed with IRAC.  
Over-densities with smaller values of $\zeta$ occurred in
only 28 of 50000 randomized catalogs, so we conclude with
much greater than $99$\% confidence that this spike is inconsistent
with a random distribution of galaxy redshifts.  
The shape of our selection function, also shown in figure~\ref{fig:zhist}, implies that
we should expect an average of roughly 1.9 galaxies 
in this redshift interval.
Our observed number, 17, therefore corresponds to
a galaxy over-density in redshift space
of $\delta_g^z \simeq 6.9\pm 2.1$ if we assume
Poisson uncertainties and neglect
the two galaxies whose redshifts defined the cluster boundary
(Steidel et al. 1998).

\subsection{Environmental Dependence of Galaxy Properties}

Shapley \et (2005) present the full population synthesis modeling of
the broad-band spectral energy distributions from rest-frame far-UV
to the rest-frame near-IR for the full sample of 72 galaxies considered
here, resulting in the most robust measurements of stellar population
parameters to date for star-forming galaxies at $z \simgt 2$.  Using these results, 
we search for possible differences in the galaxy properties  
within and outside of the identified over-density.
Figure~\ref{fig:mass_age_z} shows the inferred stellar masses and ages for
the 72 galaxies in our sample as a function of redshift, taken from \citet{shapley2005}.  
The inferred ages of galaxies within the ``spike'' clearly stand out above the rest,
with a much larger proportion of old ages compared to those outside the 
over-density. 
For constant star formation (CSF) models, the average galaxy age within the spike
is $1400\pm 180$ Myr, compared to
$720\pm103$ Myr in the ``field''\footnote{By ``field'' we mean simply the 55 galaxies
in the spectroscopic sample that lie outside of the redshift range $z=2.300\pm0.015$ containing
the ``spike''. }, which if taken literally implies average ``formation'' redshifts
of $z_f=4.3$ and $z_f=3.0$, respectively.  A 2-sample Kolmogorov-Smirnov
test on the CSF model ages yields a probability of only $1\times 10^{-4}$ that the age {\it distributions}
are the same inside and outside of the ``spike''.
As discussed extensively in \citet{shapley2005} in the context of the current
galaxy sample, the luminosity-weighted age is not a quantity which is well-determined
from population synthesis models, 
particularly for star-forming galaxies when the ``current'' star formation rate 
is high compared to the past average. It is generally possible to make the inferred age
of a galaxy significantly younger by assuming an exponentially declining star formation
rate with an e-folding time $\tau$.
For the best-fit $\tau$ model applied to each galaxy in the sample, 
the average galaxy ages are $790\pm180$ Myr inside the spike
and $440\pm90$ Myr outside, and the K-S test returns a probability of
$7.3\times 10^{-3}$ that the age distributions are the same.
Thus, while the true star formation histories of the galaxies both within and
outside of the over-density are almost certainly more complicated than the
assumed models, and the ages of individual galaxies remain uncertain \citep{shapley2005},
there is a clear difference in the age-sensitive spectral signatures between the
two sub-samples in the sense that proto-cluster galaxies are significantly older.

The differences in age are accompanied by significant differences
in stellar mass, which are much less dependent on the assumed
star formation histories (right-hand panel of figure~\ref{fig:mass_age_z})\footnote{As
discussed extensively in \citet{shapley2005}, there remains the possibility of systematic
underestimates of total stellar mass whose maximum amplitude depends on the observed  
galaxy color (under-estimates could be larger for bluer galaxies). In principle 
it is possible that the stellar masses are comparable inside and outside
the over-density, but the observed differences instead reflect the level of dominance of the current episode
of star formation over the integral of past star formation. However, since the UV luminosities and
inferred star formation rates are similar inside and outside the ``spike'', this interpretation 
would require more extreme differences in star formation history than the more plausible scenario that
both the stellar masses and ages differ.}  
with the bluest galaxies 
Both the CSF and $\tau$ fits imply that galaxies inside the spike
have mean stellar masses roughly twice as large as those outside.
A 2-sample K-S test rules out a similar stellar mass distribution
at the $\sim 2$\% level.  
In contrast, as we will shortly show,
the {\it total} masses of galaxies inside and outside
the spike should be roughly the same.  
We can therefore conclude that galaxies inside the spike have converted
a much larger fraction of their baryons into stars than
have those outside, presumably because they have had more time to do so. 
A naive interpretation of figure~\ref{fig:mass_age_z} suggests that a typical 
``field'' galaxy will reach the same level of maturity at $z\simeq 1.5$ that
has been attained by a proto-cluster galaxy at $z \sim 2.3$. 

\subsection{Theoretical Expectations}

A generic expectation of popular galaxy formation models is
that galaxies inside large-scale over-densities
should be older than those outside,
since galaxy-scale matter fluctuations inside the over-densities
are sitting on a large-scale pedestal and can more easily
cross the threshold $\delta_c=1.69$ for collapse.
The theoretically expected difference in galaxy ages can be crudely
estimated as follows.  
Given the estimate $b\sim 2.1$ (Adelberger et al. 2004)
for the galaxy bias in the BX sample, we use the approach
of \citet{steidel1998} to estimate the true mass over-density
$\delta_m$. This quantity is related to the observed redshift-space
over-density $\delta_g^z$ through the expression $1+b\delta_m=\left|C\right|(1+\delta_g^z)$,
where $\left|C\right|\equiv V_{app}/V_{true}$ is an estimate of the effects of 
redshift-space distortions caused by peculiar velocities, which is itself a function
of both $\delta_m$ and $z$. In the present case, $V_{app} \simeq 7200$ Mpc$^{3}$ is 
the co-moving volume in which the
measurement is being made, bounded by the 8\minpoint5 by
8\minpoint5 region on the plane of the sky and the co-moving distance, 
neglecting peculiar velocities, between $z=2.285$ and $z=2.315$. A crude 
approximation for the volume correction factor (see \citealt{steidel1998})
is given by 
$C=1+f-f(1+\delta_m)^{1/3}$ where $f=\Omega_m(z)^{0.6}$,
which we take to be $f=0.96$ at $z \simeq 2.3$. 
From these equations, one obtains $\delta_m\simeq 1.8$
for $\delta_g^z=6.9$, and $C=0.61$. To evaluate the linear over-density relevant
to the rest of the discussion, 
we use the approximation for spherical collapse from \citet{mo1996} (their equation 18),
which for $\delta_m=1.8$ gives $\delta_L \simeq 0.8$
as the linear matter over-density in real space of the $z=2.300$ spike.

If we imagine, for simplicity, that star formation begins at initial virialization,
the $t_{\rm sf}\sim 800$ Myr inferred age for field
galaxies at $z_0=2.3$ (for constant star formation models; \citet{shapley2005}) 
implies that their typical virialization redshift was
$z_f=3.1$.  These galaxies therefore have 
a linear over-density at redshift $z_0=2.3$ of 
$\delta_0 \sim [ (1+z_f)/(1+z_0) ]\delta_c \sim 2.1$.
The same galaxies would have linear over-densities
of $\delta_0+\delta_L\sim 2.9$ if they were within the spike.
This corresponds to a virialization redshift $z_f$ for spike
galaxies of $(1+z_f)/(1+z_0)=2.9/\delta_c$, or $z_f=4.7$.  
The implied typical age of spike galaxies would be expected to be $\sim 1600$ Myr,
or roughly twice the age of field galaxies.  If the typical age
of field galaxies is half as large, i.e., $t_{\rm sf}=400$ Myr,
the same arguments imply an age for spike galaxies of $\sim 1400$ Myr.
We would therefore expect galaxies within the spike to be roughly 2--3 times
older than those outside. As discussed above, the data support these
naive expectations. 

Our claim above that the total masses of the galaxies should be similar within and outside
the over-density follows from
the extended Press-Schechter formalism.
If $N(M|\delta_m)$ represents the Press-Schechter mass function
in a large region with (Eulerian) over-density $\delta_m$, then
$N(M|\delta_m)$
is related to the standard Press-Schechter mass function $N(M|0)$ via
\begin{equation}
N(M|\delta_m) = N(M|0)\biggl[\biggl(1+\frac{\nu^2-1}{\delta_c}\biggr)\delta_m+1\biggr]
\label{eq:mwmassfn}
\end{equation}
where $\delta_c\simeq 1.69$,
$\nu\equiv \delta_c/\sigma(M)$, and $\sigma(M)$ is the r.m.s. fluctuation
when the linear density field is smoothed on mass-scale $M$
(e.g., \citealt{mo1996}; see their equations~19 and~20).
If we assume that the galaxies in our
survey are associated with halos with dark mass $M_{\rm thresh}>10^{12}M_\odot$
\citep{adelberger2004b}, 
the expected mean halo mass inside and outside
the spike differ by only $\sim 16$\% for $\delta_m\simeq 1.8$.

\subsection{The Fate of the Spike}

We turn now to the question of what the spike will have become
by redshift $z\sim 0$.  Evolved forward in an $\Omega_M=0.3$, 
$\Omega_\Lambda=0.7$ cosmology, its linear over-density of $\delta_L\sim 0.8$ at $z=2.3$
corresponds to a linear over-density of $\delta_L\sim 2$ at $z=0$.
Since this exceeds the collapse threshold $\delta_c=1.69$, we expect
the spike to have virialized by $z=0$.  The mass of the virialized
object is easy to calculate.  The approach of \citet{steidel1998}
implies that the spike is associated with an Eulerian matter
over-density of $\delta_m \simeq 1.8$ at $z=2.3$, so its mass is
$(1+\delta_m)\bar\rho V_{true}\sim 1.4\times 10^{15} M_\odot$
where $\bar\rho$ is the mean co-moving matter density and 
$V_{true}\sim 12000$ Mpc$^3$ is approximately the volume
within the observed over-density, corrected for the effects of peculiar velocities
using the factor $C$ from above. 
We conclude that the spike is likely to evolve into a relatively massive 
cluster by $z\sim 0$.  Our observations therefore imply that at least
some clusters already contained old galaxies with large stellar masses by $z\sim 2.3$, 
in good agreement with inferences based on studies of cluster ellipticals 
at zero-to-moderate redshifts.

\subsection{Proto-Cluster Completeness}

Our census of galaxies within the proto-cluster is far from complete.
Because the spectroscopic completeness of BX/MD galaxy candidates detected to
$K_s \simeq 22$ is only $\sim 24$\% within the $\sim 70$ arcmin$^2$ region considered
here, and the galaxies in the spectroscopic sample have been observed without
regard to their near-IR properties,  we expect that the total number of ``spike'' galaxies with
properties similar to those observed and within the same solid angle is $\simgt 70$.  
It is difficult to estimate the
number of galaxies that are within the ``spike'' but not selected by our rest-UV
color criteria; crude estimates of the completeness with respect to all relatively
massive galaxies present at $z \simgt 2$ suggest that there could be another factor
of $\simlt 2$ in galaxies that have similar masses to the ones that are selected \citep{shapley2005,
reddy2005a}. Redshift estimates based on photometry, even when the IRAC bands are included,
are too uncertain ($\sigma_z \simeq 0.3$) to assign galaxies to a particular 
structure in redshift space, and spectroscopy will be difficult or 
impossible for galaxies with no current star
formation at $z \sim 2.3$ using current-generation telescopes and instruments. It is possible
that there could be a significant number of massive but quiescent red galaxies associated with
the ``spike'' that would make the differences between the proto-cluster and ``field''
environments still more significant at $z \simeq 2.300$. 

The angular scale of the observed field was set (rather arbitrarily) by the size
of the available field with the WIRC near-IR imager and LRIS. 
As shown in figure~\ref{fig:field_map},
the ``spike'' objects are not obviously distributed differently on the plane of the
sky than the ``field'' objects--a 2-D, 2-sample K-S test gives a probability of 17\% that the
proto-cluster and ``field'' galaxy sky distributions are the same. 
There is no evidence that the observed region has
included the whole structure, or even that it is roughly centered
on the structure, although the angular extent of the observed field (corresponding
to a co-moving transverse scale of $\sim 14$ Mpc at $z=2.300$) is roughly
the co-moving scale expected for a proto-cluster.  The current observational situation for
the HS1700+64 $z=2.300$ over-density is similar
to earlier observations of the $z=3.09$ structure in the SSA22 field 
\citep{steidel1998,steidel2000},
where the field size for the initial discovery was also $\sim 9$\arcm\, and
where a similar redshift-space over-density in both LBGs and 
Lyman $\alpha$ emitters was found. 
Subsequent analysis of narrow-band Lyman $\alpha$ images
over a much larger field by \citet{hayashino2004} has shown that the initial SSA22 pointing
was indeed the densest proto-cluster-sized sub-region within a large-scale structure
extending over a $\sim 60$ Mpc co-moving scale.

The $z=2.300$ structure, which (as for the higher-redshift SSA22 example) was discovered 
serendipitously within a redshift survey, might also be compared to more 
``pointed'' observations surrounding
high-redshift radio galaxies. Perhaps most comparable is PKS 1138$-$262 ($z=2.16$), which
has been studied using narrow-band Lyman $\alpha$ imaging and spectroscopy
\citep{kurk2000,pentericci2000}, narrow-band
H$\alpha$ imaging \citep{kurk2004a}, and broad-band near-IR photometry \citep{kurk2004b}.
While a considerable number of objects associated with the radio galaxy have been
identified, the pointed nature of the observations makes an evaluation of the significance
of the numbers difficult to evaluate; \citet{kurk2004b} estimate the over-density of 
Lyman $\alpha$ emitters in redshift space to be somewhat smaller than that found in the
SSA22 field at $z=3.09$ by \citet{steidel2000}, $\delta_g^z \sim 4.4$ evaluated
over a somewhat smaller field but a very similar redshift range. An evaluation
of the implied mass scale of the 1138$-$262 structure\footnote{There is an error in \citet{kurk2004b}
in calculating the value of $\bar\rho V$, which for their assumed cosmology
should be $2.0\times 10^{14}$ M$_{\sun}$, not $6.6\times10^{15}$ M$_{\sun}$.}, calculated in
a manner consistent with 
that used above for the $z=2.300$ HS1700$+$64 proto-cluster, yields $M_{\rm tot}\sim 6\times 10^{14}$
M$_{\sun}$ for the PKS 1138$-$262 structure, with an associated linear matter
over-density of $\delta_L \simeq 0.6$.   

To date, the $z=3.09$ structure in SSA22 and the $z=2.30$ structure in the HS1700+643
field are the most statistically significant cluster-scale over-densities 
known at redshifts $z \simgt 1.5$.
While the implied over-densities may be somewhat larger than in the radio galaxy
fields, the real difference is that identification of the structures within a 
controlled redshift survey provides a more direct means of measuring the
redshift-space over-density and estimating the true matter over-density. These
quantities require knowledge of both the mean density of objects that are being used
as a ``tracer'', and their bias with respect to matter fluctuations.
The more general surveys  
also offer the possibility of examining galaxy properties over the
full dynamic range of environment, from proto-cluster to proto-void.

\section{Summary and Discussion}
\label{sec:discussion}

We have shown that the HS1700+643 field contains a highly significant over-density 
($\delta \sim 7$ in redshift space) of galaxies at $z=2.300\pm0.015$. A crude analysis
of this structure indicates that it has a mass scale of $\sim 1.4\times10^{15}$ M\subsun
and that it will become a rich cluster of galaxies by $z\sim0$. While it is not the
highest redshift ``proto-cluster'' identified to date, its significance is more
easily established because the redshift survey in which it was discovered provides
the context necessary for estimating the associated matter over-density. 
The presence of the ``proto-cluster'' embedded within the redshift survey 
allows us to make the first direct assessment of
the differences in galaxy properties as a function of environment at high redshift,
using mid-IR observations from {\it Spitzer}/IRAC together with extensive ground-based
data.  Galaxies within the proto--cluster have mean inferred 
stellar masses a factor
of $\sim 2$ times higher and mean inferred ages $\sim 2$ times older than
similarly-UV-selected galaxies in the ``field''.  
This lends support to a simple theoretical picture in which the main difference
between dense environments and the ``field'' is an earlier virialization time
and hence a larger stellar mass fraction at a given cosmic time. 

The differences in inferred age between the proto-cluster and ``field'' environment,
while highly significant at $z=2.30$, would be relatively subtle by $z \sim 0$, 
where (average) age
differences of $\simlt 1$ Gyr relative to ages in the range $11-12$ Gyr would 
probably be ``in the noise'' of spectroscopic
or photometric age dating for local early-type galaxies (e.g., \citealt{hogg2004}).  
Another complication
in comparing the high redshift results with local ones is that even galaxies inhabiting
environments that we call the ``field'' at $z=2.30$ may have descendants
that find themselves in relatively dense environments (even if they are not in clusters)
by $z \sim 0$ (e.g., \citealt{baugh1998,adelberger2004b}); here it would be 
important to quantify the local density for 
``field'' ellipticals in order to reconcile the high-redshift results with
those based on the ``fossil record''. It would require 
a substantial leap of faith to connect the $z \sim 2.3$ galaxies with 
lower-redshift descendants of a particular morphology (e.g., ``early type galaxies''); 
however, the ages and stellar masses of the
UV-selected denizens of the HS1700+643 protocluster seem  
consistent both with expectations based on the empirical fossil record and with 
the generic expectations of hierarchical structure formation.

The results presented above show that, at least at $z \simgt 2$, a significant fraction
of objects destined for rich cluster environments are still forming stars,  and that
selection in the rest-UV, which is largely independent of past star formation
\citep{shapley2005},
is an effective and efficient means of assembling samples of galaxies that allow
quantitative assessment of the dependence of galaxy properties on environment. 
Structures like the $z=2.300$ proto-cluster in the HS1700+643 field offer many
opportunities for focused follow-up observations aimed at better understanding
the physics of galaxy formation within cluster environments; in particular, $z=2.300$
is a redshift at which both narrow-band Lyman $\alpha$ (observed wavelength of 4012 \AA)
and H$\alpha$ (2.166$\mu$m) are easily accessible from the ground, where a
combination of rest-frame optical and rest-frame far-UV spectroscopy offers
a wealth of information on the kinematics, chemistry, and stellar populations 
of cluster galaxies in the pre-cluster environment, and where the impact of galaxy
formation on the proto-intracluster medium may be measured. The results of such
ongoing follow-up studies will be presented elsewhere.\\

We are indebted to the IRAC instrument team, particularly Pauline Barmby, Giovanni Fazio, Jiasheng Huang,
and Peter Eisenhardt, for obtaining and reducing the Spitzer/IRAC data in the HS1700+643
field, and for allowing us early access to the data. We thank the referee, Paul Francis, for
a constructive review of the original version of the paper. 
CCS, DKE, and NAR have been supported by grant
AST03-07263 from the
US National Science Foundation and by the David and Lucile Packard Foundation.  
KLA acknowledges support
from the Carnegie Institution of Washington, and AES from the Miller Institute for
Basic Research in Science.

\bibliographystyle{apj}

\end{document}